\documentclass[a4paper,twocolumn]{article}

\usepackage{float}       
\usepackage{helvet}      
\usepackage{rsc}         
\usepackage{setspace}    
\usepackage{amsmath,amssymb}
\usepackage[usenames,dvipsnames]{color}
\usepackage{psfrag,graphicx}

\newcommand{\etal}{{\it et al.} }

\renewcommand\mark[1]{\bgroup\color{red}\bfseries{[#1]}\egroup}
\newcommand\remove[1]{\bgroup\color{Gray}\bfseries{[#1]}\egroup}

\title{{Fluctuations and correlations during the shear flow of elastic particles
  near the jamming transition}}

\author{Claus Heussinger, Pinaki Chaudhuri and Jean-Louis Barrat\thanks{Universit\'e Lyon 1, Laboratoire de Physique de la Mati\`ere
  Condens\'ee et Nanostructures, CNRS, UMR 5586
  Domaine Scientifique de la Doua F-69622 Villeurbanne Cedex, France}
}

\begin{document}
\maketitle
\bibliographystyle{rsc}

\begin{abstract}
  We present a numerical study of the flow of an assembly of frictionless,
  {soft discs} at zero temperature, in the vicinity of and slightly above
  the jamming density. We find that some of the flow properties, such as the
  fluctuations in the number of contacts or the shear modulus, display a
  critical like behaviour that is governed by the proximity to the jamming
  point. Dynamical correlations during a quasistatic deformation, however, are
  non critical and dominated by system size.  At finite strain rates, these
  dynamical correlations acquire a finite, strain-rate dependent amplitude, that
  decreases when approaching the jamming point from above.

  {\bf Manuscript submitted to the themed Issue on granular and jammed materials of Soft Matter}
\end{abstract}



\section{Introduction  and background}
The properties of granular materials, in particular sand, are a constant source
of fascination for children and adults alike, and are intrinsically related to
the ability of such systems to exist in either solid or fluid states, under very
similar conditions.  For the scientist this fascination may arise through the
apparent contradiction between the rigidity of the individual grains and the
fragility of the assembly as a whole. This means that, for example, small
changes in the loading conditions (such as changing the inclination angle of the
support) can lead to large scale structural rearrangements (``avalanches'') or
even to the complete fluidization of the material.  A few years ago, Liu and
Nagel ~\cite{liuNATURE1998} have suggested a ``phase-diagram'' for this type of
solid-liquid transition (``jamming'').  At zero temperature the axes relate to
the ways an unjamming transition can be triggered, either by increasing the
external driving (e.g.  shear stress) or by decreasing the density of the
material. The present study will probe the vicinity of this "unjamming line"
using quasistatic and finite strain rate simulations of a model granular system.

If one applies a shear stress, which is small and below a certain
  threshold (``yield-stress''), the material will respond as an elastic solid.
Increasing the stress above the yield-stress, the particles will unjam and start
to flow. This flow behavior is called ``plastic flow'' as the material will not
revert to its original shape when the stress is removed.

In the following we will assume that the system can flow at arbitrary small
strain-rates without showing flow-localization. That this is possible is by no
means guaranteed, as in some instances coexistence of flowing and jammed states
is observed ~\cite{denninJPhysCondMatt2008}. We have
not observed such a persistent strain localization in the simulations that are
presented here.

Plastic flow is observed in a large number of glassy materials, that are a
priori very different from the athermal granular systems close to point J (see
below) studied in this paper.  However, all materials display a rather universal
behavior, that was illustrated already in early studies on the plastic-flow of
metallic glasses~\cite{argonACTA1979,argonACTA1983}. These studies have given
indications that in the flowing phase the main plastic activity is spatially
localized to so called shear transformation
zones~\cite{falklangerPRE1998,bouchbinderPRE2007a}. These zones are non
persistent, localized in space, and presumably consist of a few atoms that
undergo the irreversible rearrangements responsible for the observed plastic
flow.  Recently, this plasticity has been further analyzed in simulations with a
focus on the quasistatic dynamics at small strain rates, close to the flow
  arrest~\cite{malandroJCP1999,MaloneyPRE2006,lemaitreCaroliPRE2007,tanguyEPJE2006,tsamadosEPJE2008,tsamadosPRE2009}.
With these studies it was possible to trace back the origin of plastic activity
to the softening of a vibrational mode and the vanishing of the associated
frequency~\cite{malandroJCP1999}. In real space, this softening is associated
with the formation of distinct, localized zones where the plastic failure is
nucleated~\cite{MaloneyPRE2006}. In turn, this can trigger the failure of nearby
zones, such that avalanches of plastic activity form that may
``propagate'' through the entire system.


It has been argued that the macroscopic extent of these avalanches is a
signature of the quasistatic dynamics, which gives the system enough time to
propagate the failure throughout the system. Beyond the quasistatic regime,
i.e. farther away from the jammed state,  the size of these events is expected to
be finite. Thus, one naturally finds an increasing length-scale that is
connected with the flow arrest upon reducing the stress towards the threshold
value~\cite{bocquetPRL22009,PicardPRE2005,lemaitrePRL2009}.

Without external drive, an (un)jamming transition can occur for decreasing
particle volume fraction below a critical value, $\phi_c$. This special point,
which is only present in systems with purely repulsive steric interactions, has
been given the name ``point J''~\cite{ohern03,majmudarPRL2007}. At this point
the average number of particle contacts jumps from a finite value $z_0$ to zero
just below the transition. The value of $z_0$ is given by Maxwell's estimate for
the rigidity transition \cite{maxwell1864,calladine78} and signals the fact that
at point J each particle has just enough contacts for a rigid/solid state to
exist. This marginally rigid state is called ``isostatic''. Compressing the
system above its isostatic state a number of non-trivial scaling properties
emerge~\cite{ohern03,durianPRL1995}. As the volume fraction is increased,
additional contacts are generated according to $\delta\!
z\sim\delta\!\phi^{1/2}$.  The shear modululs scales as $G\sim p/\delta\!  z$
and vanishes at the transition (unlike the bulk modulus)\cite{ohern03}.  This
scaling is seen to be a consequence of the non-affine deformation response of
the system~\cite{wyart05c}, with particles preferring to rotate around rather
than to press into each other~\cite{EllenbroekPRL2006}. Associated with the
breakdown of rigidity at point J is the length-scale, $l^\star\sim \delta
z^{-1}$~\cite{wyart05b,wyart05a}, which quantifies the size over which
additional contacts stabilize the marginally rigid isostatic state.

In this article we present results from quasistatic and small strain-rate flow
simulations {of a two-dimensional system} in the vicinity of point J.
Together with the linear elastic shear modulus, at point J also the yield-stress
$\sigma_y$
vanishes~\cite{peyneauPRE2008,olssonPRL2007,hatanoJPSJ2008,heussingerPRL2009,otsukiPRE2009,xuPRE2006}.
Thus, point J is connected with a transition from plastic-flow behavior
($\phi>\phi_c$, $\sigma_y>0$) to normal fluid flow ($\phi<\phi_c$,
$\sigma_y=0$), with either Newtonian~\cite{olssonPRL2007} or Bagnold
rheology~\cite{hatanoJPSJ2008,otsukiPRE2009} at small strain-rates.  In
consequence, both (un)jamming mechanisms as described above are present at the
same time: the flow arrest, as experienced by lowering the stress towards
threshold, is combined with the vanishing of the threshold itself.

In this study we want to address two questions: In how far do the general
plastic flow properties carry over to this situation of small or, indeed,
vanishing yield-stress? Is the vicinity to point J and its isostatic state at
all relevant for the flow properties ? {It will be shown that while the
  stress fluctuations reflect the critical properties at point J, dynamical
  correlations are typical those observed in the flow of elasto-plastic solids}.

We will approach these questions starting with the quasistatic-flow regime. The
advantage of quasistatic simulations is to provide a clean way of accessing the
transition region between elastic, solid-like behavior and the onset of flow. In
the quasistatic regime flow is generated by a succession of (force-)equilibrated
solid states. Thus, one can connect a liquid-like flow with the ensemble of
solid states that are visited along the trajectory through phase-space.

In Section~\ref{sec:qs-sim}, we study the instantaneous statistical properties
of the configurations generated by this flow trajectory at zero strain rate, and
show that they display large fluctuations in several quantities, that are
associated with the proximity to the jamming point.

In Section~\ref{sec:chi4}, we follow the analysis of recent experiments
\cite{lechenault} and use a "four point correlation" tool to define a dynamical
correlation length that characterizes the extension of the dynamical
heterogeneities observed in the flow process. This dynamical length scale is
shown to scale as the system size in the zero strain rate limit, independently
of the distance to point J. The heterogeneity in the system is maximal for
strains that correspond to the typical duration between the plastic avalanches
described above.

We complement this analysis with preliminary results from dissipative
molecular-dynamics simulations that access strain-rates above the quasistatic
regime. This allows us to assess the importance of dynamic effects in limiting
access to certain regions of the landscape. Indeed, the results at larger strain
rate are system size independent, and reveal a surprising growth of the strength
of the heterogeneities with increasing packing fraction away from $\phi_c$.

\section{Simulations}

Our system consists of $N$ soft spherical particles with harmonic contact
interactions
\begin{equation}\label{eq:harmInteraction}
E(r) = k(r-r_c)^2\,.
\end{equation}
Two particles, having radii $r_i$ and $r_j$, only interact, when they are ``in
contact'', i.e. when their distance $r$ is less than the interaction diameter
$r_c=r_i+r_j$.  This system has been studied in several contexts, for example
in~\cite{ohern03,olssonPRL2007,haxtonPRL2007}.  The mixture consists of two
types of particles ($50:50$) with radii $r_1=0.5d$ and $r_2=0.7d$ in
two-dimensions. Three different system-sizes have been simulated with $N=900$,
$1600$ and $2500$ particles, respectively. The unit of length is the diameter,
$d$, of the smaller particle, the unit of energy is $kd^2$, where $k$ is the
spring constant of the interaction potential.  We use quasistatic shear
simulation, and compare some of the results with those obtained from dissipative
molecular-dynamics simulations at zero temperature.

Quasistatic simulations consists of successively applying small steps of shear
followed by a minimization of the total potential energy. The shear is
implemented with Lee-Edwards boundary conditions with an elementary strain step
of $\Delta\gamma=5\cdot10^{-5}$. After each change in boundary conditions the
particles are moved affinely to define the starting configuration for the
minimization, which is performed using conjugate gradient
techniques~\cite{lammps}. The minimization is stopped when the nearest energy
minimum is found. Thus, as the energy landscape evolves under shear the system
always remains at a local energy minimum, characterized by a potential energy, a
pressure $p$ and a shear stress $\sigma$.

The molecular dynamics simulations were performed by integrating Newton's
equations of motion with elastic forces as deduced from
Eq.~(\ref{eq:harmInteraction}) and dissipative forces
\begin{equation}\label{eq:friction}
{\vec F}_{ij} =  - b\left[\left( {\vec v}_i - {\vec v_j}\right)\cdot {\hat r}_{ij}\right]{\hat r}_{ij},
\end{equation}
proportional to the velocity differences along the direction ${\hat r}_{ij}$
that connects the particle pair. The damping coefficient is chosen to be $b=1$.
Rough boundaries are used during the shear, the boundaries being built by
freezing some particles at the extreme ends in the $y$-direction, from a
quenched liquid configuration at a given $\phi$. The system is sheared by
driving one of the walls at a fixed velocity in the $x$ direction, using
periodic boundary conditions in this direction.

For all system sizes, the distance between the top and bottom boundaries is
$52.8d$ and each of the boundaries has a thickness of $4.2d$. The system size is
changed by modifying the length of the box in the $x$-direction.

\section{Results}

\subsection{Quasistatic simulations}\label{sec:qs-sim}

\begin{figure}[h]
 \begin{center}
   \includegraphics[width=0.4\columnwidth,angle=-90]{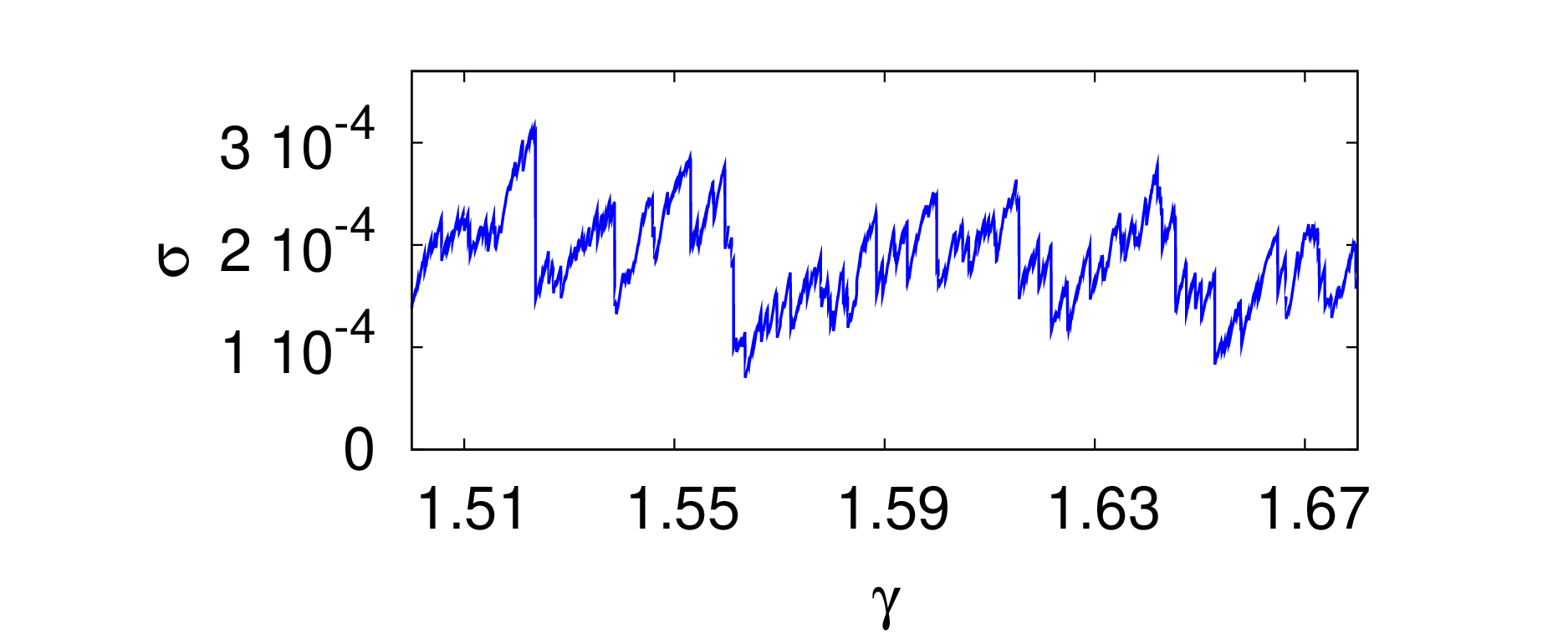}
   \hfill
   \includegraphics[width=0.4\columnwidth,angle=-90]{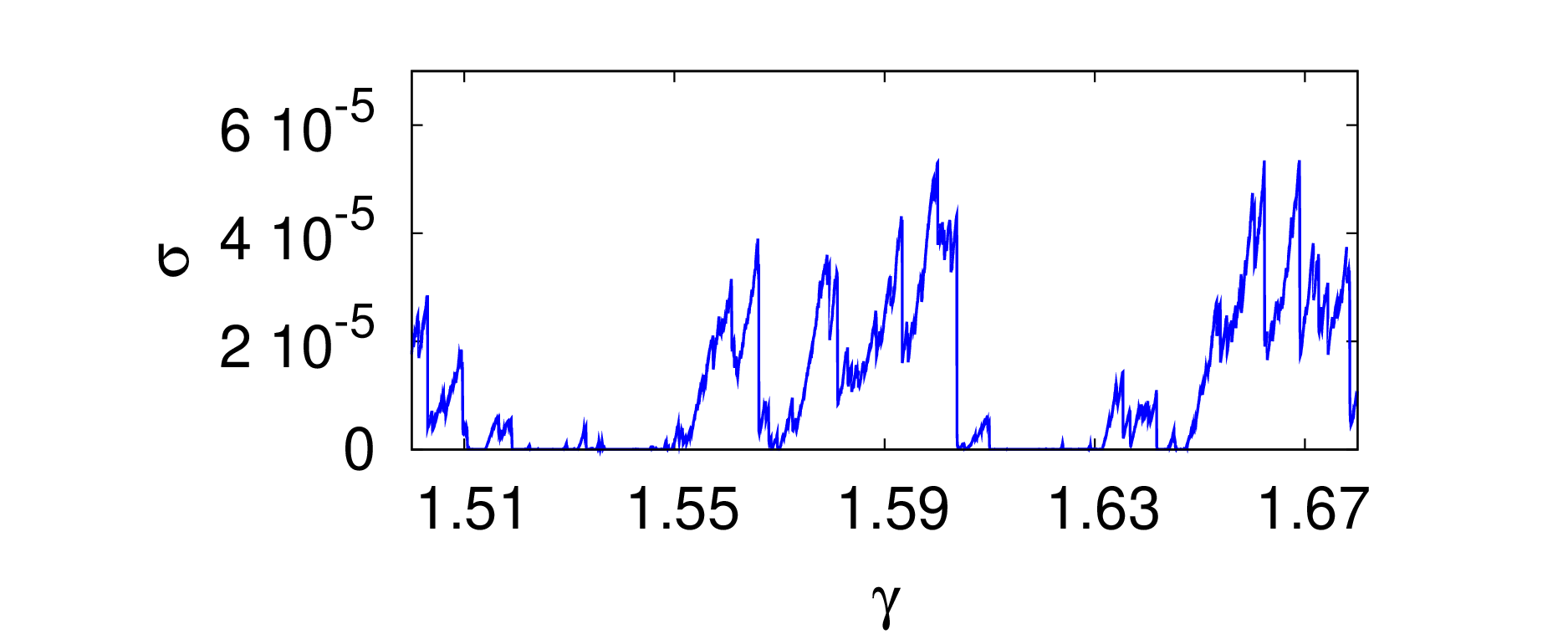}
\end{center}
\caption{Stress-strain relation for two different volume-fractions, $\phi=0.847$
  (top) and $\phi=\phi_c=0.8433$ (bottom). At volume-fractions close to $\phi_c$
  the signal is intermittent showing long quiet regions where the system flows
  without the building up of stress.}\label{fig:stress_strain}
\end{figure}

As is readily apparent from Fig.~\ref{fig:stress_strain}, a typical feature of
quasistatic stress-strain relations is the interplay of ``elastic branches'' and
``plastic events''.  During elastic branches stress grows linearly with strain
and the response is reversible. In plastic events the stress drops rapidly and
energy is dissipated.

{In setting the elementary strain step, $\Delta\gamma$, care must be taken to
properly resolve these events. Too large strain steps would make the simulations
miss certain relaxation events. We chose a strain step small enough, such that
most minimization steps do not involve any plastic relaxations.  In consequence,
the elastic branches are well resolved, each consisting of many individual
strain steps.}

\begin{figure}[h]
 \begin{center}
   \includegraphics[width=0.49\columnwidth]{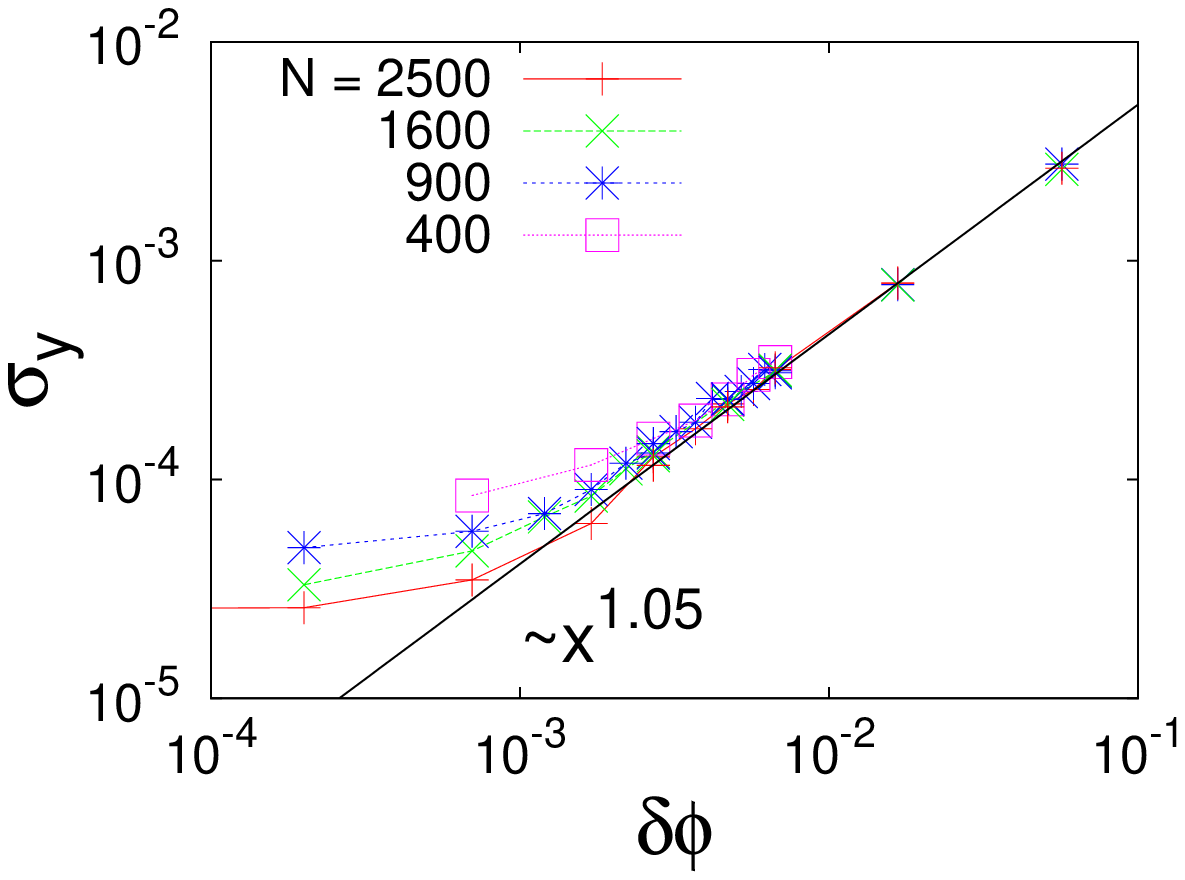}
   \hfill
   \includegraphics[width=0.49\columnwidth]{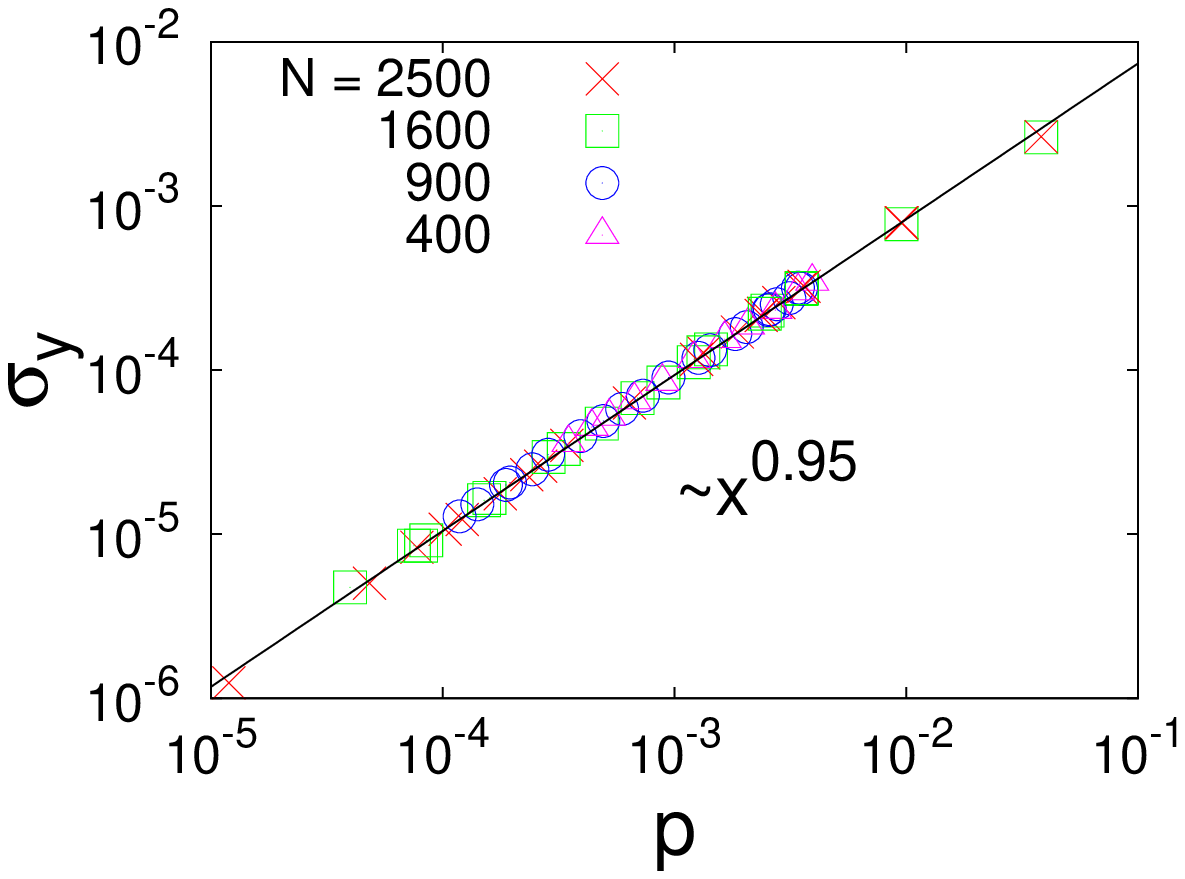}\end{center}
 \caption{Average yield-stress as function of volume-fraction (left) and
   pressure (right). The yield-stress is determined as an average over
   stress-values just before plastic events occur, i.e at the top end of each
   elastic branch.}\label{fig:yieldstress}
\end{figure}

The succession of elastic and plastic events defines the flow of the material
just above its yield-stress $\sigma_y(\phi)$. The value of the yield-stress
depends on volume-fraction and nominally vanishes at $\phi_c$ (see
Fig.~\ref{fig:yieldstress}). For finite systems, however, finite-size effects
dominate close to $\phi_c$ such that one cannot observe a clear vanishing of
$\sigma_y$.  Rather, as Fig.~\ref{fig:stress_strain} shows, one enters an
intermittent regime, {i.e. a finite interval in volume-fraction in which}
the stress-signal shows a coexistence between jammed and ``freely-flowing''
states. This is evidence of a distribution of jamming thresholds, $P(\phi_c)$,
which sharpens with increasing the system-size~\cite{ohern03,heussingerPRL2009}.
A finite-size scaling analysis of this distribution allows one to extract the
critical volume-fraction. {To this end we count the number of jamming
  events that lead from the freely-flowing state to the jammed state and
  back~\footnote{For a closer illustration of a jamming event see the supporting
    material to our previous paper~\cite{prl2009supplmat}}. We find a maximum
  number of events at a certain $\phi_c(L)$, which can be extrapolated to
  $L=\infty$ to define the critical volume fraction of our simulation.} The
value we find, $\phi_c=0.8433$ is slightly higher than what has been obtained
previously, however, evidence is mounting that $\phi_c$ is
non-universal~\cite{chaudhuriCM2009} and depends on the details of the ensemble
preparation. Scaling properties in the vicinity of a jamming threshold, on the
other hand, appear to be universal~\cite{chaudhuriCM2009}.

In Fig.~\ref{fig:yieldstress} we display the yield-stress $\sigma_y$, as a
function of volume-fraction, $\delta\phi=\phi -\phi_c$, and pressure $p$.
Finite-size effects are particularly strong when using $\phi$ as control
variable. In the intermittent regime the average stress levels off to a
system-size dependent value.

Much better scaling behavior can be obtained, when using pressure as control
variable, as this is characterized by the same finite-size effects as the
shear-stress. In the following we will therefore use pressure as control
variable.  Be aware, however, that we do \emph{not} run pressure-controlled
simulations, as for example Peyneau and Roux~\cite{peyneauPRE2008} but use the
average pressure, $\langle p \rangle(\phi)$, only to plot our simulation
results. The value $\sigma/p\approx0.1$ obtained from Fig.~\ref{fig:yieldstress}
is consistent with these pressure-controlled simulations. On the other hand, the
scaling with volume-fraction, $p\sim \delta\phi^{1.1}$, is slightly stronger
than in linear elasticity at zero stress~\cite{ohern03,durianPRL1995}, where the
pressure simply scales as $\delta\phi$.  In view of  the strong finite-size effects,
the scaling with volume-fraction should, however, be taken with care.

In the following we show results from five different volume-fractions,
$\phi=0.846$, $0.848$, $0.85$, $0.86$ and $\phi=0.9$, which are all above
$\phi_c$ and \emph{outside} the intermittent regime {(i.e. no
  freely-flowing zero-stress states occur)}. For each volume-fraction we study
three different system sizes with $N=900,1600$ and $2500$ particles.

\subsubsection{Elastic properties}

As reviewed in the introduction a hallmark of the elasticity of solids in the
vicinity of point J is the scaling of the linear elastic shear modulus $g\sim
p^{1/2}$. Similarly, the number of inter-particle contacts scale as
$z=z_0+Ap^{1/2}$.

\begin{figure}[h]
 \begin{center}
   \includegraphics[width=0.6\columnwidth,angle=-90]{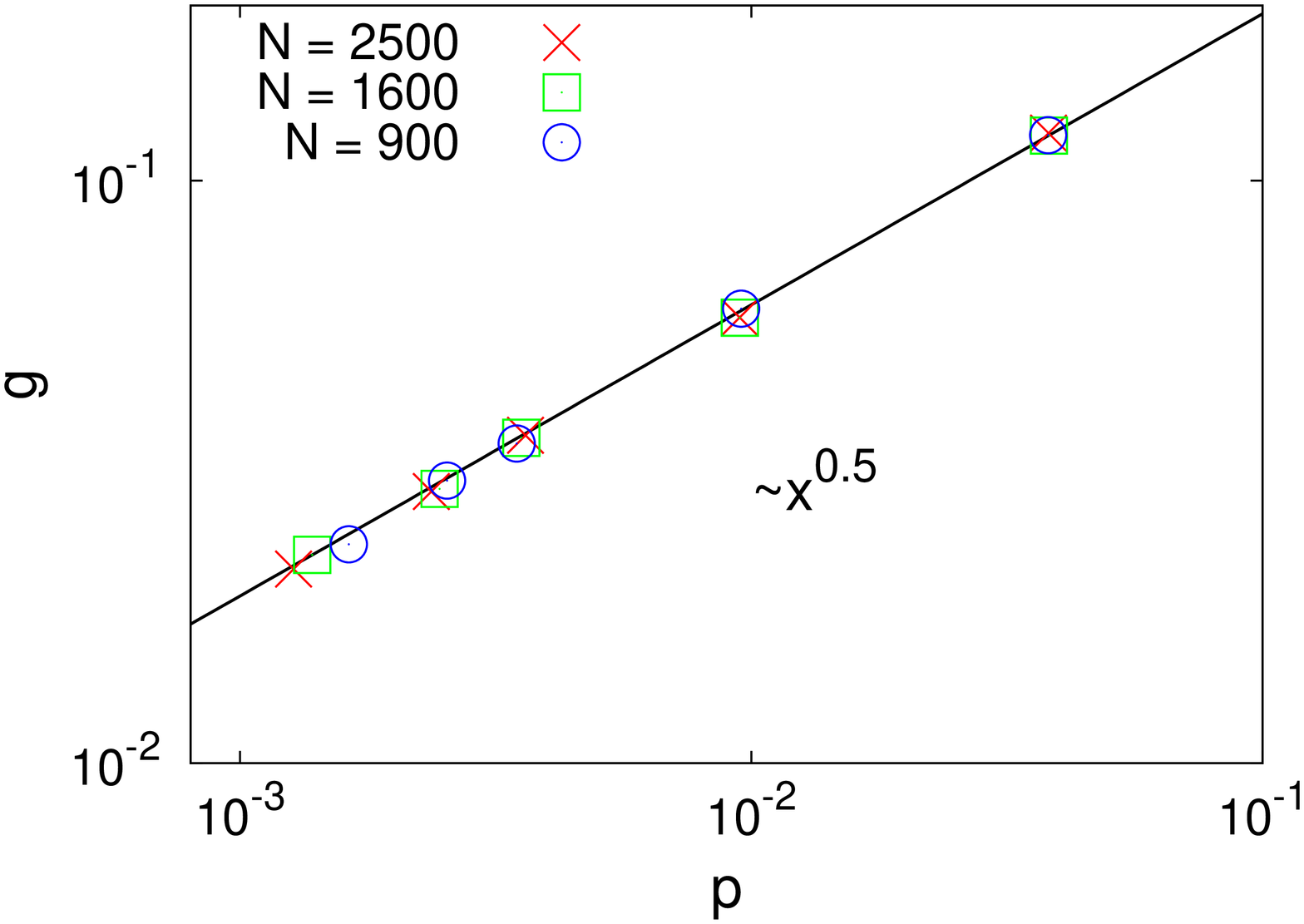}
   \includegraphics[width=0.6\columnwidth,angle=-90]{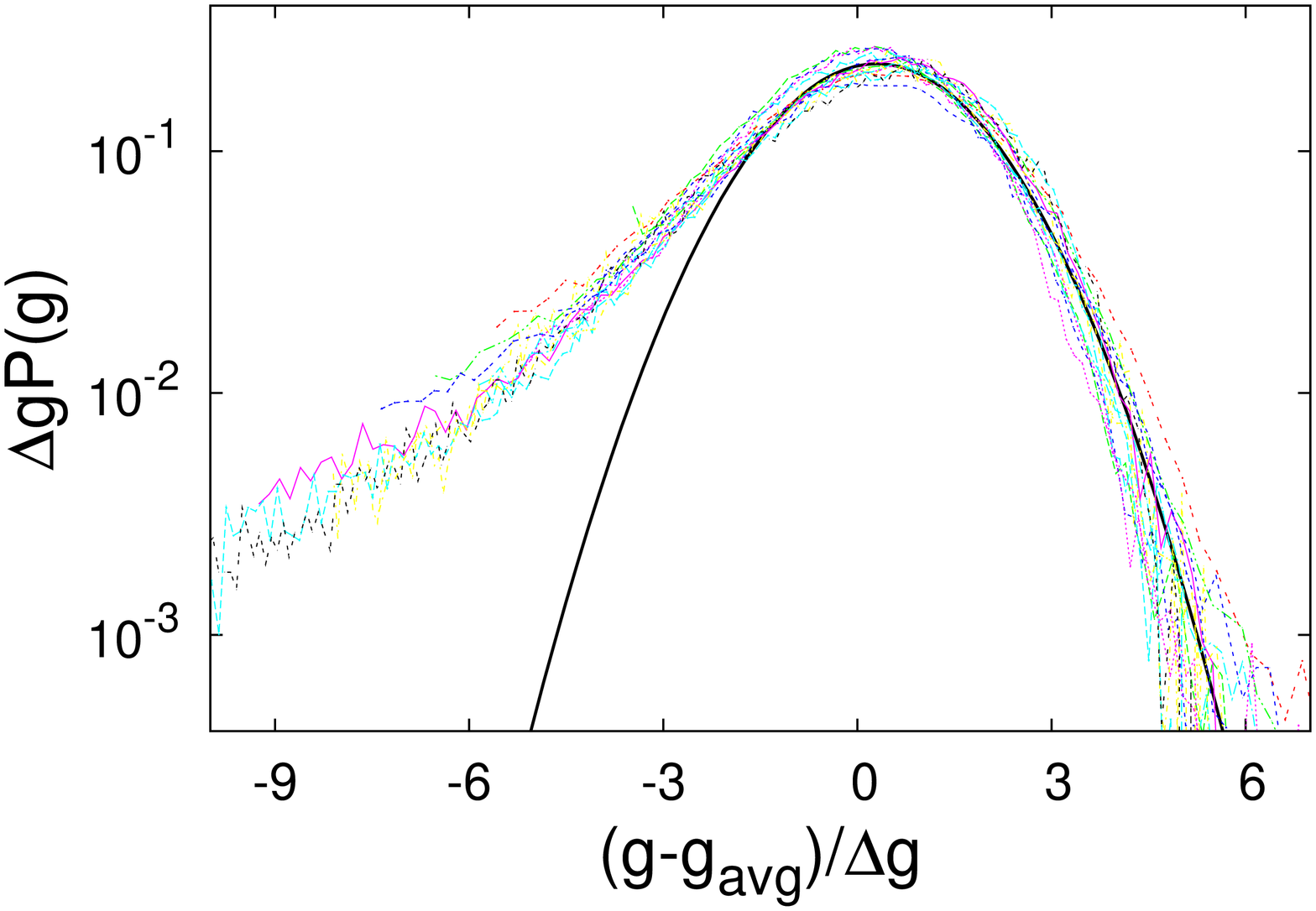}
\end{center}
\caption{(Top) Average nonlinear elastic shear modulus $g_{\rm avg}$ as function
  of pressure $p$. (Bottom) Probability distribution $P(g)$ centered around
  average value $g_{\rm avg}$ and rescaled width according to $\Delta g=
  p^{0.25}/N^{0.5}$. Black solid line is a Gaussian pdf. }\label{fig:g.p}
\end{figure}

\begin{figure}[h]
 \begin{center}
   \includegraphics[width=0.6\columnwidth,angle=-90]{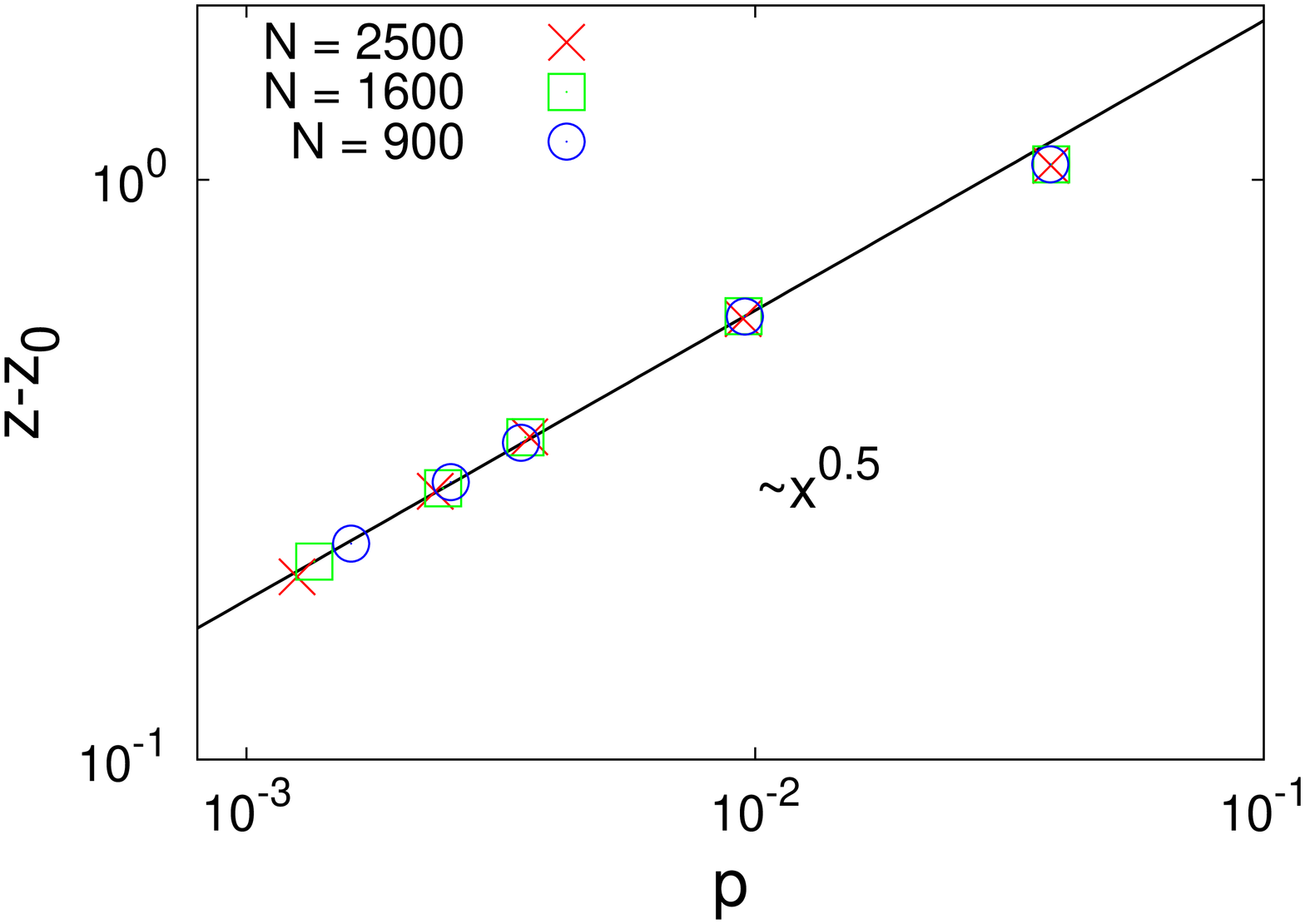}
   \hfill
   \includegraphics[width=0.6\columnwidth,angle=-90]{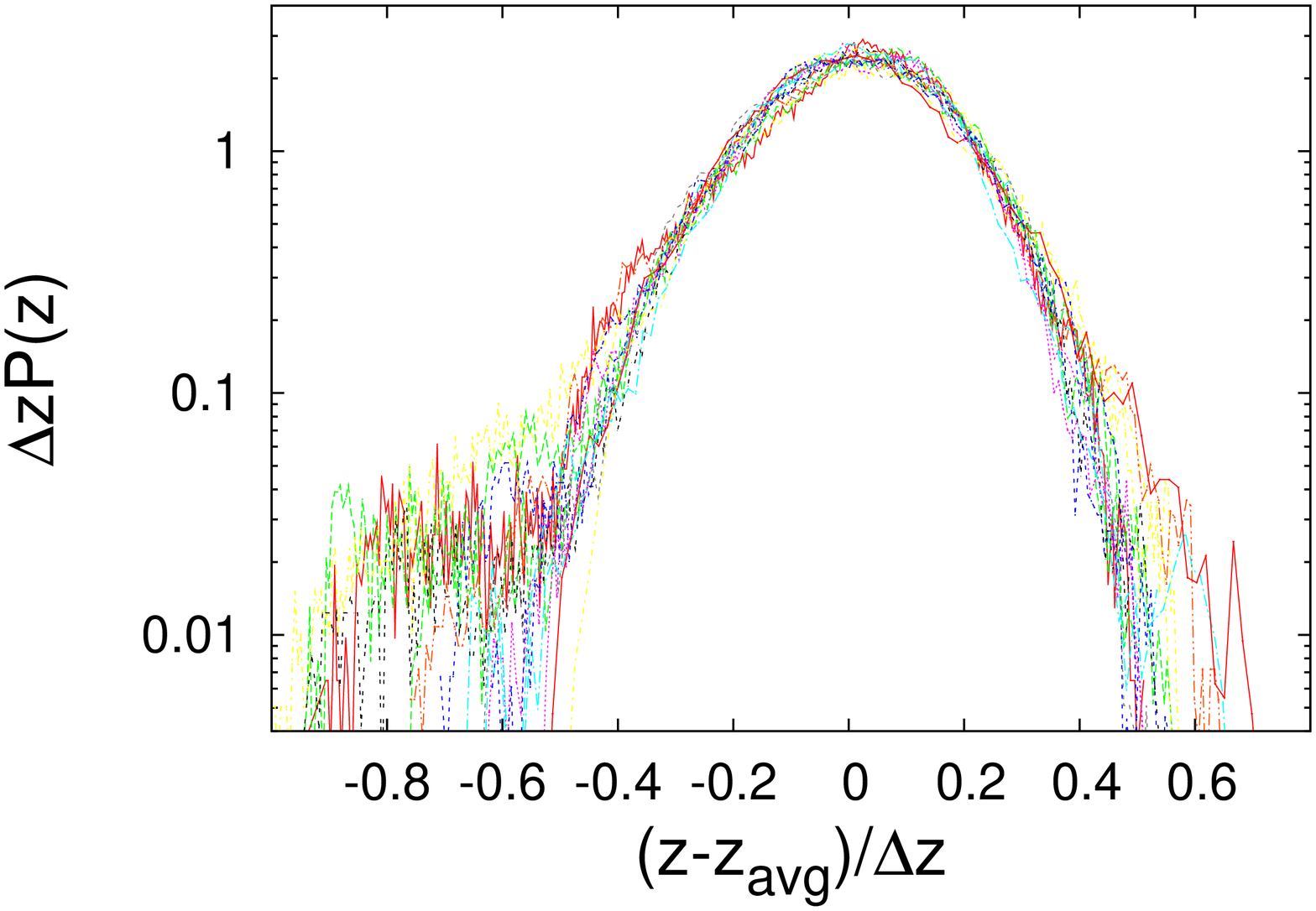}
\end{center}
\caption{(Top) Average contact number $z$ as function of pressure $p$. The
  values for $z_0\equiv z(p=0)$ are determined from a best fit. They are smaller
  than $z_{c}=4$ as to the presence of rattlers, which have not been accounted
  for. (Bottom) Probability distribution $P(z)$ centered around average value
  $z_{\rm avg}$ and rescaled width according to $\Delta z=
  p^{-0.35}/N^{0.5}$}\label{fig:z.p}
\end{figure}

We have analyzed the elastic branches in the steady-state flow to find
(Figs.~\ref{fig:g.p} and \ref{fig:z.p}) that the same scaling properties
characterize the average nonlinear elastic modulus $g_{\rm avg}$, which we
define as the local slope of the stress-strain curve, and also the associated
contact numbers $z_{\rm avg}$. If we take these scaling properties as a
signature of the criticality of point J, we can conclude that for the range of
volume-fractions considered we are in the ``critical regime''.

As additional characterization of the ensemble of elastic states we report the
probability distributions of shear moduli and contact numbers, respectively.
Maybe surprisingly all the obtained distributions have approximately the same
shape and can be superimposed on a single master curve. To achieve this we
center each distribution around the average value and rescale the width with a
factor $p^\alpha N^\beta$.

By looking carefully at the individual distributions we do observe a slight
trend towards the development of non-Gaussian tails close to $\phi_c$.  While
non-Gaussian distributions are to be expected close to critical
points~\cite{binderZPhysikB1981}, the effect is quite small and all
distributions have a well developed Gaussian core.  The pronounced small-$g$
tail of $P(g)$ is due to shear moduli that extend down to zero. Similar tails
have been observed in \cite{MaloneyPRE2006} and related to a softening of the
response upon approach towards plastic instabilities.  Indeed, we found that
manually suppressing states close to plastic events, the weight in the small-$g$
tail is reduced.



For the width of the g-distribution we obtain $\Delta g = p^{0.25}/N^{0.5}$.
Thus, the absolute width of the distribution decreases with decreasing pressure,
while the relative width, $\Delta g/g_{\rm avg}$ diverges at point J. For the
contact numbers, on the other hand, we find a divergence of the absolute width
itself, $\Delta z = p^{-0.35}/N^{0.5} $. These enhanced fluctuations certainly
support the view of $\delta z$ as an order parameter for a continuous jamming
transition. The quantity $\Delta z^2N$ would then be analogous to a
susceptibility, $\chi\sim p^{-\gamma}$, diverging with an exponent $\gamma=0.7$.

Our results are different than those of Henkes and
Chakraborty~\cite{henkesPRE2009}, where fluctuations of $z$ are found to be
independent of pressure, $\Delta z\sim p^0$.  Note, however, the subtle
difference in ensemble. These authors study a pressure-ensemble, {in which
  states are generated by quenching \emph{random} particle configurations to the
  local minimum of the potential energy landscape (similar to the procedure in
  Ref.~\cite{ohern03}). One may view these states as the inherent structures of
  a high temperature liquid.  Our ensemble then corresponds to the inherent
  structures of a driven glassy material. We fix volume-fraction and sample only
  states that are connected by the trajectory of the system in phase-space. The
  ensemble therefore reflects the dynamics of the system and the region of
  phase-space where it is guided to.}

\subsubsection{Yield properties}

We now go beyond the properties of the elastic states and discuss aspects
related to their failure during the plastic events. As indicated in the
introduction, plastic events can be viewed as bifurcations in energy-landscape.
A local energy minimum vanishes and the system has to search for a new minimum
at lower energy and stress. In quasistatic dynamics this process is
instantaneous. The associated stress-drop is therefore visible as a vertical
line in the stress-strain relation (Fig.~\ref{fig:stress_strain}).

\begin{figure}[t]
 \begin{center}
   \includegraphics[width=0.8\columnwidth,angle=0]{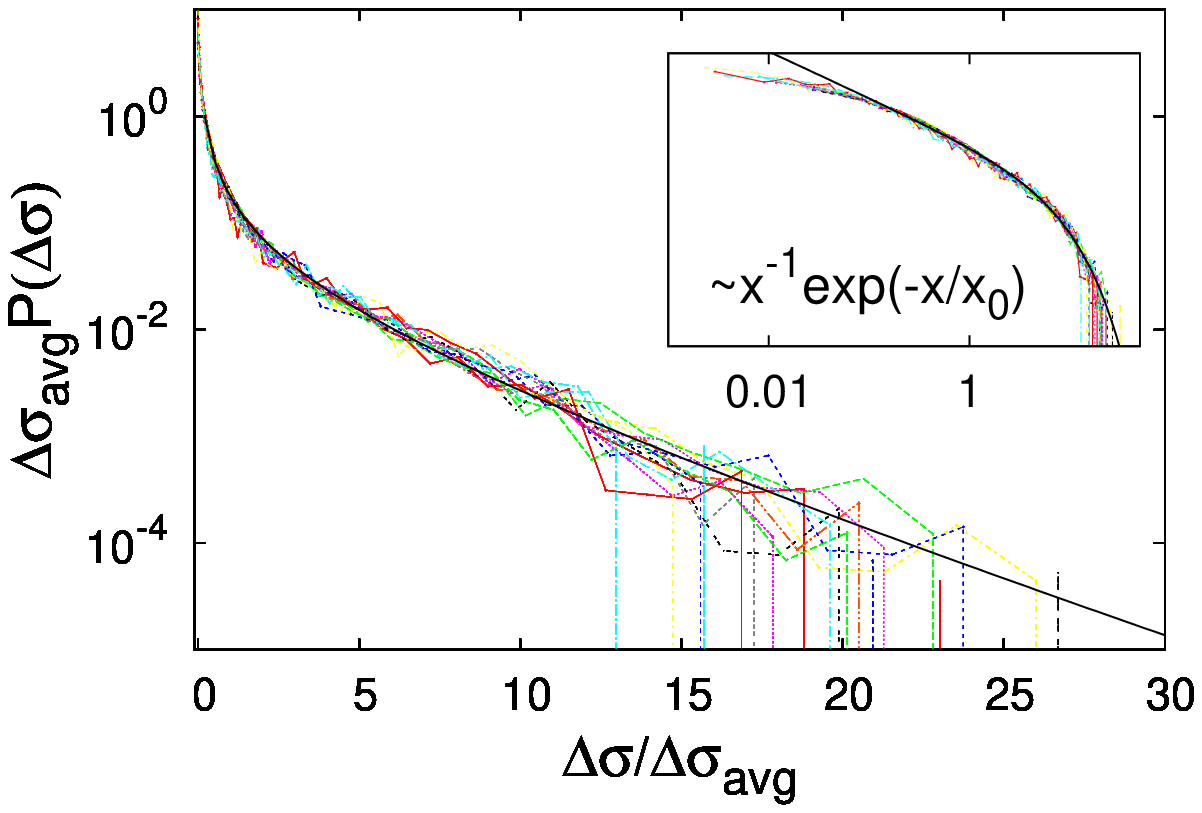}
   \hfill
   \includegraphics[width=0.6\columnwidth,angle=-90]{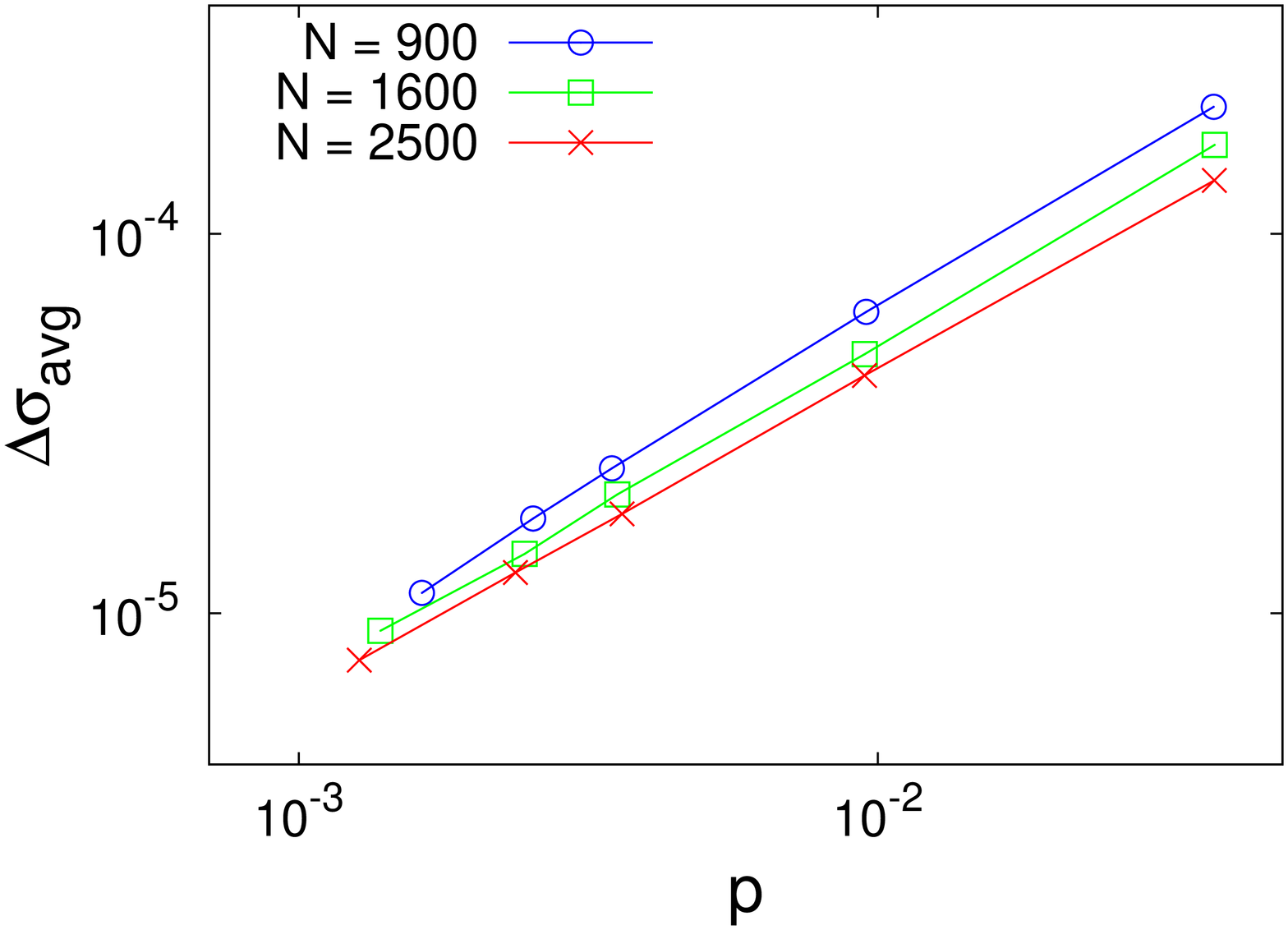}
\end{center}
\caption{(Top) Distribution of stress-drops normalized with average values
  $\Delta\sigma_{\rm avg}$. Inset shows the same figure in a log-log
  representation.  (Bottom) Scaling of $\Delta\sigma_{\rm avg}$ with pressure
  $p$. 
}\label{fig:distrDrops}
\end{figure}

\begin{figure}[h]
 \begin{center}
   \includegraphics[width=0.6\columnwidth,angle=-90]{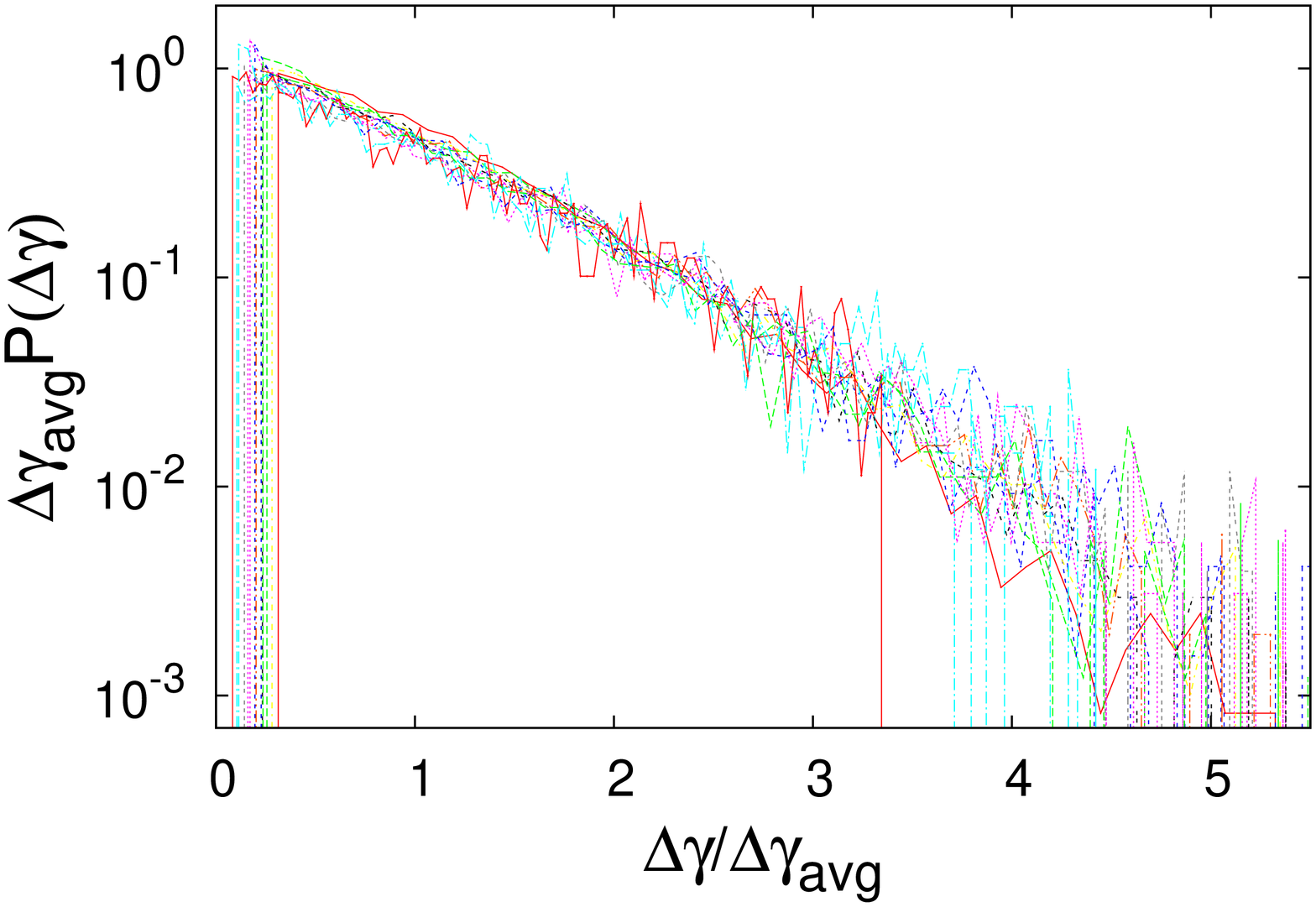}
   \hfill
   \includegraphics[width=0.6\columnwidth,angle=-90]{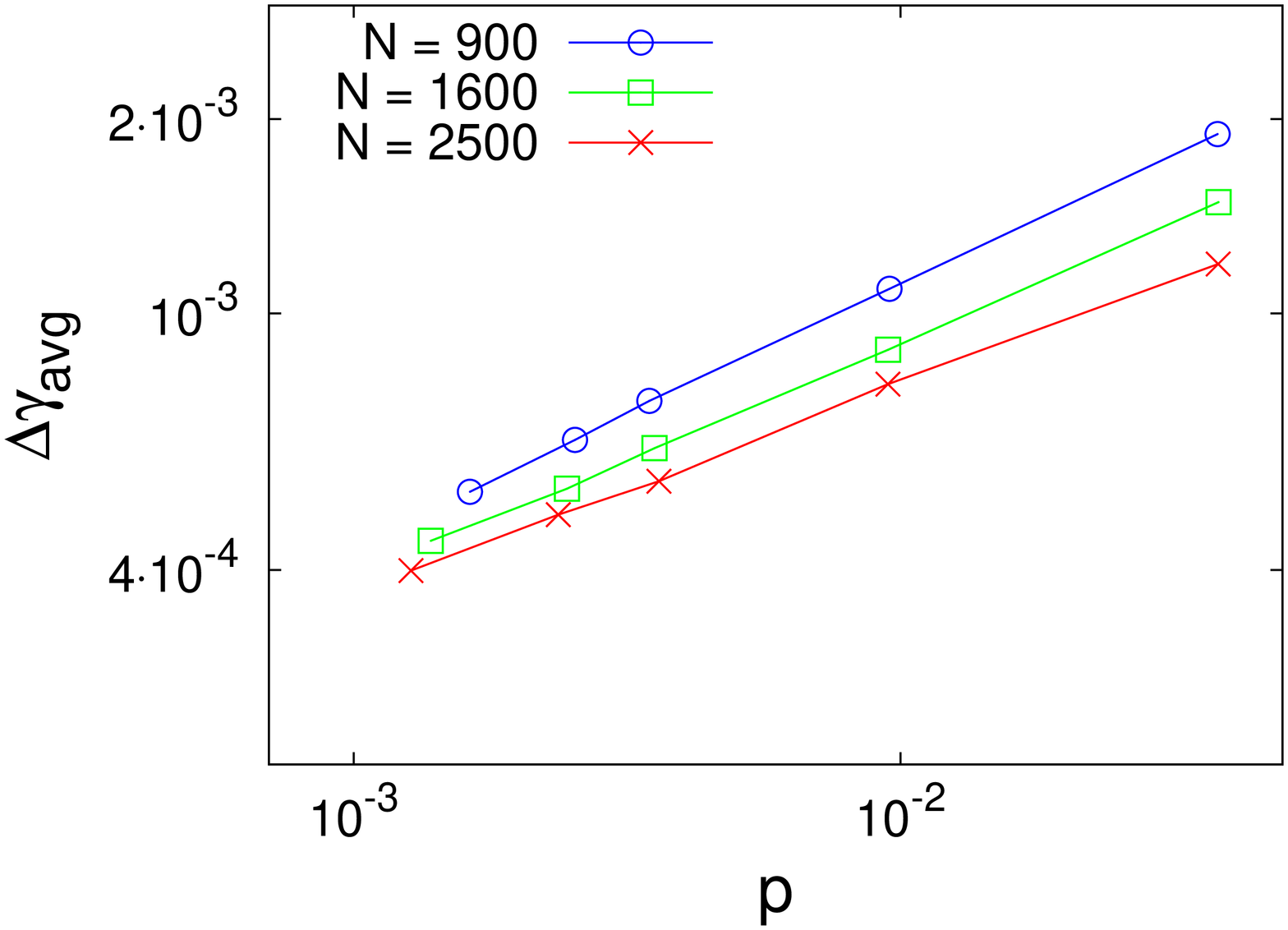}
\end{center}
\caption{(Top) Distribution of elastic branch length normalized with average
  values $\Delta\gamma_{\rm avg}$. (Bottom) Scaling of $\Delta\gamma_{\rm avg}$
  with pressure $p$. 
  The ratio $g=\Delta\sigma_{\rm avg}/\Delta\gamma_{\rm avg}$, which defines a
  shear-modulus, is consistent with the scaling in Fig.~\ref{fig:g.p}.
}\label{fig:elBranches}
\end{figure}
In the following we characterize the amount of dissipation during a plastic
event by the associated stress-drop, $\Delta\sigma$. The frequency of plastic
events is discussed in terms of the length of elastic branches,
$\Delta\gamma_{\rm avg}$.

Just like the probability distributions within the elastic states, the functions
$P(\Delta\sigma)$ and $P(\Delta\gamma_{\rm avg})$ (Figs.~\ref{fig:distrDrops} and
\ref{fig:elBranches}) are universal and can be rescaled on a single
master-curve. Here, it is sufficient to use the first moment of the
distribution, i.e. the ensemble-averaged stress drops and elastic-branch
lengths, respectively. As the logarithmic scale in the inset in
Fig.~\ref{fig:distrDrops} shows, the collapse for the stress-drop distribution
is quite good for large as well as for small stress drops. The black line
represents a fit of the form $\Delta\sigma^{-1}\exp(-\Delta\sigma/\sigma_L)$
with the stress-scale $\sigma_L\approx 5\Delta\sigma_{\rm avg}$. The
intermediate power-law behaviour $P\sim\Delta\sigma^{-1}$ reflects the lack of
scale related to a typical event size. The only relevant scale is the
exponential cut-off at $\sigma_L$. Tewari \etal\cite{tewariPRE1999} have
reported an exponent of $-0.7$ in the energy-drop distribution at finite-strain
rates. The simulated systems are somewhat smaller, however. Kabla \etal
\cite{kablaJFM2007} have found an exponent of $-1.5$ in a vertex model for
foams, in agreement with renormalization group arguments~\cite{dahmenPRL2009}.

The exponential tail has been observed in several different studies
\cite{tsamadosEPJE2008,tsamadosEPJE2008,lernerPRE2009,MaloneyPRE2006,baileyPRL2007}
in two and in three spatial dimensions.  Tsamados \etal\cite{tsamadosEPJE2008}
have furthermore related this feature of the stress-drop distribution to the
diversity of local flow-defects causing the plastic event.

A similar universality has been observed by Maloney and
Lema\^{i}tre~\cite{MaloneyPRE2006}. Their simulations are conducted with three
different interaction potentials but without changing the density, which is set
to high values far away from the rigidity transition. The authors have argued
for a universal value of the ``flow-strain'' $\sigma_y/g$ of a few percent.
Apparently, this can only be true far away from $\phi_c$. As the yield-stress
vanishes faster than the shear modulus, one finds a ratio $\sigma_y/g\sim
\delta\phi^{1/2}$ that vanishes at point J. Thus, particle configurations at the
onset of jamming are highly fragile and susceptible to even minute changes in
the boundary conditions.

The average stress-drop as well as the average length of elastic branches change
with pressure and system-size as displayed in Figs.~\ref{fig:distrDrops} and
\ref{fig:elBranches} ~\footnote{Note, that these values also depend on
    the elementary strain step used in the simulations. For any finite step-size
    there will be some small plastic events, that cannot be resolved but only
    lead to an apparent reduction of the stress increase.  This will
    artificially increase the length of elastic branches and decrease the weight of the small
    $\Delta\sigma$-tail of the stress-drop distribution.}. As a function of
pressure one observes an increase but with a slope that depends on system-size.
The average stress-drops increase somewhat slower with pressure than the
yield-stress~\footnote{ The effective exponents range from $0.8$ to $0.9$ as
  compared to an exponent ${0.95}$ for the yield-stress}. The relative stress
fluctuations $\Delta\sigma_{\rm avg}/\sigma_y$ are thus slightly enhanced at
small pressures close to $\phi_c$. The same trend is also visible in the total
stress fluctuations as calculated by $\left\langle (\sigma-\langle
  \sigma\rangle)^2\right \rangle$.

The overall scale of both, $\Delta\sigma_{\rm avg}$ and $\Delta\gamma_{\rm
  avg}$, decreases with system-size to give a smooth stress-strain relation in
the thermodynamic limit. Previous
studies~\cite{tanguyEPJE2006,MaloneyPRE2006,tsamadosEPJE2008} have observed a
scaling of the stress-drops with $N^{-1/2}$. This includes~\cite{MaloneyPRE2006}
a system of harmonically interacting particles, as studied here, but at a rather
high pressure. In general, we observe a weaker dependence on system-size, with
an effective exponent that increases with pressure. Our data is consistent,
however, with the value of $1/2$ being the relevant high-pressure limit.

\subsection{Dynamical correlations}
\label{sec:chi4}

Let us now turn to the dynamics of the system. In particular we want to
characterize dynamic correlations in the motion of particles. While at
volume-fractions above $\phi_c$ the isostaticity length-scale $l^\star$ is
clearly finite~\footnote{If we assume for the isostaticity length
  $l^\star=1/\delta z$, we would have values $l^\star \approx5$ at $\phi=0.846$
  and $l^\star\approx 1$ at $\phi=0.9$ ($z$-values taken from
  Fig.~\ref{fig:z.p}).}, there is nevertheless a large dynamical length-scale
related to the flow arrest. This has, for example, been evidenced in a system of
Lennard-Jones particles with dissipative dynamics~\cite{lemaitrePRL2009}. We
will show below that a similar length-scale occurs in our system of purely
repulsively interacting particles, independent of the distance to $\phi_c$.

Let us start by presenting the results from the quasistatic simulations.  To
define a dynamical correlation length we study heterogeneities in the particle
mobilities. To this end we use the
overlap-function~\cite{glotzerJCP2000,franzJPhysCM2000}
\begin{equation}\label{eq:Q_definition}
  \langle Q(\gamma,a)\rangle =\left\langle \frac{1}{N}\sum_{i=1}^N\exp\left[-\frac{u_{i\rm
          na}(\gamma)^2}{2a^2}\right] \right\rangle\,,
\end{equation}
of particles undergoing nonaffine displacements $u_{i\rm na}$ during a strain
interval of $\gamma$~\footnote{{To calculate the overlap function we use
    the non-affine displacements in the gradient direction (irrespective of the
    distance to the wall).}}. Particles moving farther than the distance $a$
(``mobile''), have $Q\approx 0$, while those that stay within this distance
(``immobile'') have $Q\approx 1$.

As a function of strain $\gamma$, the average
overlap $\langle Q\rangle$ will decay, when particle displacements $u_{\rm na}$
are comparable to the probing length-scale $a$. The overlap function is similar
to the intermediate scattering function with wave-vector $q\sim 1/a$. Thus, $a$
sets the probing length-scale. The decay of $Q(\gamma,a)$ then gives an
associated structural relaxation strain, $\gamma^\star(a)$, on which particle
positions decorrelate.

\begin{figure}[h]
 \begin{center}
   \includegraphics[width=0.8\columnwidth]{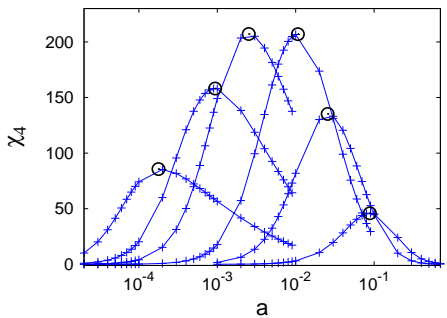}
\end{center}
\caption{Dynamical susceptibility $\chi_4$ as function of probing length-scale
  $a$ for various different strains $\gamma=(5,20,50,200,500,2000)\cdot 10^{-5}$
  (from left to right) and $\phi=0.9$. The maxima of the curves (black circles)
  define the amplitude $h(\gamma)$.}\label{fig:chi4}
\end{figure}

In the following we are interested in the dynamical heterogeneity of $Q$ and the
fluctuations around its average value
\begin{equation}\label{eq:chi4}
  \chi_4(a,\gamma) = N\left(\left\langle Q(\gamma,a)^2\right\rangle - \left\langle
    Q_a(\gamma,a)\right\rangle^2\right)\,,
\end{equation}
which defines the (self-part of the) dynamical susceptibility $\chi_4$. This is
displayed in Fig.~\ref{fig:chi4} as a function of both strain $\gamma$ and
probing length-scale $a$. For each $\gamma$ it has a well defined peak (at
$a^\star(\gamma)$) that parallels the decay of the overlap function $\langle
Q\rangle$~\cite{lechenault}.

The strength of the correlations are encoded in the peak-height,
$h(\gamma)\equiv\chi_4(a^\star(\gamma),\gamma)$ (black circles in
Fig.~\ref{fig:chi4})). As $\chi_4$ can be written as the integral over a
correlation function, it is connected to the correlation volume, or to the
number of correlated particles.  Assuming that this volume forms a compact
region in space~\cite{dauchot2005PRL,droccoPRL2005} we can relate the amplitude
of $\chi_4$ to a dynamic correlation length via $\xi^2(\gamma)= h(\gamma)$.

\begin{figure}[t]
 \begin{center}
   \includegraphics[width=0.8\columnwidth]{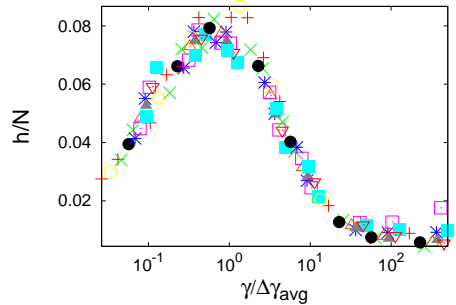}
\end{center}
\caption{Amplitude $h(\gamma)$ as taken from the quasistatic simulations. Data
  for different volume-fractions and system-sizes. The axes are normalized
  according to the scaling form $h(\gamma)=N\tilde h(\gamma/\Delta\gamma_{\rm
    avg})$, with $\Delta\gamma_{\rm avg}$ taken from
  Fig~\ref{fig:elBranches}.}\label{fig:chi4_amplitude}
\end{figure}

Following the maxima in Fig.~\ref{fig:chi4} from left to right, one sees that
the amplitude first increases and then quickly drops to small values. This
implies that there is a finite strain $\gamma$, at which $\chi_4$ presents an
\emph{absolute} maximum. To extract this maximum we plot in
Fig.~\ref{fig:chi4_amplitude} the amplitude $h(\gamma)$ for various
volume-fractions $\phi$ and system-sizes $N$.

There are two surprising features in this plot.

First, by rescaling the strain-axis with the average length of elastic branches,
$\Delta\gamma_{\rm avg}$ (see Fig.~\ref{fig:elBranches}) we find reasonable
scaling collapse for all studied volume-fractions and system-sizes. {This
  implies that cooperativity, as measured by the amplitude of $\chi_4$ and the
  length of elastic branches are intimately related}. The frequency of plastic
events sets the strain-scale for dynamical heterogeneities. As the length of the
elastic branches decreases with system size, the absolute maximum shifts towards
smaller strains, with $h(\gamma)$ becoming effectively a decreasing function of
$\gamma$ in the thermodynamic limit.

The second surprising feature in Fig.~\ref{fig:chi4_amplitude} is the
system-size dependence of $h$. It turns out that $h/N$ rather than $h$ itself is
independent of system-size, indicating a finite variance of the distribution of
$Q$ values in the thermodynamic limit (see Eq.~(\ref{eq:chi4})). Assuming the
connection with the correlation length to hold, $\xi^2\sim h$, this implies a
correlation length that is proportional to the length of the simulation box,
$\xi\approx 0.3L$, independent of volume-fraction and distance to point J. This
illustrates the fact that quasistatic dynamics is inherently dominated by
system-size effects, as already shown in previous works
~\cite{lemaitrePRL2009,PicardPRE2005,heussingerPRL2009,maloneyPRL2006}.
{Note, that this system-size dependence is of different origin than the
  finite-size effects present within the above mentioned intermittent regime,
  which occurs close to $\phi_c$. The intermittency can be avoided by staying
  away from $\phi_c$. In contrast, the system-size dependence encountered here
  is quite independent of volume-fraction, but rather a generic feature of the
  quasistatic regime, as we show now.}

\begin{figure}[t]
 \begin{center}
   \includegraphics[width=0.8\columnwidth]{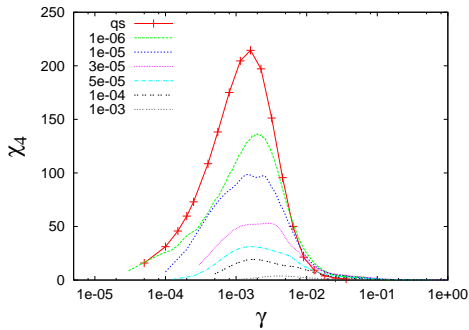}
\end{center}
\caption{$\chi_4(\gamma)$ for different strain-rates $\dot\gamma$ and $a=0.01$.
  comparison with quasistatic (`qs') simulations. Note the different boundary
  conditions used. MD simulations are with walls, while quasistatic simulations
  have periodic boundary conditions. This may explain the difference in the
  amplitude.}\label{fig:chi4_strainrate}
\end{figure}

To this end let us turn to the molecular-dynamics simulations. We will show that
the dependence on system-size indeed reflects the saturation of a length-scale
that is finite for larger strain-rates and increases towards the quasistatic
regime~\footnote{A more detailed account of these simulations will be presented
  in: P. Chaudhuri and L. Bocquet, in preparation (2010).} .  As
Fig.~\ref{fig:chi4_strainrate} shows, the amplitude of $\chi_4$ increases when
reducing the strain-rate ($a=0.01$, $\phi=0.9$) and approaches the quasistatic
limit for small strain-rates. Also the strain $\gamma_m$, at which $\chi_4$ is
maximal is very well reproduced in the dynamic simulation.

\begin{figure}[t]
 \begin{center}
   \includegraphics[width=0.95\columnwidth]{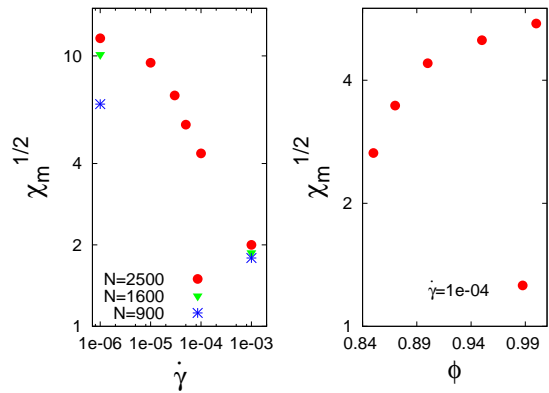}
\end{center}
\caption{(Left) Peak-height $\chi_m=\chi_4(\gamma_m)$ as determined from
  Fig.~\ref{fig:chi4_strainrate}. The saturation at small strain-rates is an
  indication of the quasistatic limit, in which $\chi_m\sim N$. In contrast, at
  high strain-rates no significant dependence on system-size is observed.
  (Right) Peak-height $\chi_m$ as function of
  volume-fraction.}\label{fig:xi_dotgamma_phi}
\end{figure}

Fig.~\ref{fig:xi_dotgamma_phi} demonstrates a saturation of the amplitude
$\chi_m\equiv \chi_4(\gamma_m)$ at small strain-rates indicating that the
quasistatic regime is entered. Comparing with the quasistatic simulation, we
find somewhat smaller values for the amplitude. It should be remembered,
however, that different boundary conditions have been used. The rough walls used
in the molecular dynamics simulations are likely responsible for the reduction
of the peak-height as compared to the quasistatic simulations (which are
performed with periodic boundary conditions).

The presence of the quasistatic regime is also evidenced by the fact that the
amplitude $\chi_m$ within the plateau depends on system-size, just as in the
quasistatic simulations. Outside this regime, on the other hand, no significant
$N$-dependence is observed. {In effect this means that the quasistatic
  regime shrinks with increasing system-size. The strainrate $\dot\gamma_{\rm
    qs}(N)$ that describes the crossover to the quasistatic regime decreases
  with $N$. This is in line with Refs.\cite{lemaitrePRL2009,PicardPRE2005},
  where a power-law dependence $\gamma_{\rm qs}\sim1/N$ is reported.  From our data
  we cannot make any definitive statement about this dependence.}

The results are furthermore consistent with those of Ono \etal\cite{onoPRE2003}.
The lowest strain-rate accessible in the latter study was $\dot\gamma=0.0001$.
At this strain-rate the correlation length was observed to be on the order of
$3$ in agreement with our data.

We also probed the volume-fraction dependence, by performing runs at
$\phi=0.85,0.87,0.9,0.95$ and $1$. The resulting amplitude of $\chi_4$ is given in
Fig.~\ref{fig:xi_dotgamma_phi}. Interestingly we observe a mild increase in the
amplitude with volume-fraction, signalling enhanced correlations \emph{away}
from $\phi_c$.

This is not due to the special choice of the parameter $a=0.01$. We have found
the same trend when fixing $\gamma$ and viewing $\chi_4$ as function of $a$ as
for the quasistatic simulations in Fig.~\ref{fig:chi4}.  Finally, we have also calculated the
\emph{absolute maximum} of $\chi_4$, viewed as function of both $a$ and
$\gamma$. In all cases, the amplitude increases with volume-fraction. 
Both, the increase of the length-scale with lowering the strain-rate and the
increase with volume-fraction are consistent with the recently proposed
elasto-plastic model of Bocquet \etal\cite{bocquetPRL22009}.


Given the trend in Fig.~\ref{fig:xi_dotgamma_phi}, one may speculate about a
vanishing dynamical correlation length (taken at constant strain-rate
$\dot\gamma=10^{-4}$ outside the quasistatic regime), as $\phi_c$ is approached.
Such a behavior is indeed compatible with our data and has recently been
observed in the rheology of a concentrated emulsion confined in gaps of
different thickness~\cite{goyonNature2008}.  {Here we interpret this
  surprising feature in the following way: at a given packing fraction,
  dynamical correlations increase with decreasing $\dot{\gamma}$, and saturate
  at a $N$-dependent value in the quasistatic regime.  This cross-over to the
  quasistatic regime does not only depend on system-size but also on packing
  fraction: $\dot{\gamma}_{\rm qs} (N,\phi )$.  
  
  For $\phi$ closer to $\phi_c$ the energy landscape becomes increasingly flat.
  Particle relaxations take longer~\cite{hatanoPRE2009} and smaller strain-rates
  are needed to allow for full relaxation into the local energy minimum.  Thus,
  smaller strain-rates are needed to reach the quasistatic behavior and
  $\dot\gamma_{\rm qs}$ decreases towards $\phi_c$.  Hence, reducing $\phi$
  towards $\phi_c$ {\it at a fixed strain rate} is "equivalent" to increasing
  the strain rate relative to $\dot{\gamma}_{\rm qs} $, and results in a
  decrease of dynamical correlations. In contrast, corrrelations taken at a
  strain rate $\dot\gamma=\dot\gamma_{\rm qs}(\phi)$, are independent of $\phi$
  and given by their quasi-static values, as displayed in
  Fig.~\ref{fig:chi4_amplitude}.}

\section{Discussion and Conclusion}

We have discussed the small strain-rate elasto-plastic flow of an athermal model
system of soft harmonic spheres. In particular, we were interested in the flow
properties at and above a critical volume-fraction (point J), at which the
yield-stress of the material vanishes. This regime combines the more traditional
elasto-plastic flow of solids above their yield-stress with the breakdown of the
rigidity of the solid state at point J.

We found that this breakdown is visible in the ensemble of states visited during
a flow simulation in a similar way as in the linear elasticity of the solid. For
example (Fig.~\ref{fig:z.p}), we showed that the average number of particle
contacts scale with the square-root of pressure, just as in linear elasticity.
In contrast, the fluctuations around this average value show a distinct behavior
that has not been observed previously. We showed that the contact-number
fluctuations actually diverge upon approaching the critical volume-fraction from
above, making the contact number an ideal candidate for an order parameter of a
continuous jamming transition as observed under steady shear. The relative
fluctuations of the shear modulus and those of the shear stress also diverge in
the same limit.
Going beyond the characterization of the average elastic properties we have
studied the statistics of plastic events (Figs.~\ref{fig:distrDrops} and
\ref{fig:elBranches}). It seems that all distributions have universal scaling
forms reminiscent of standard critical phenomena.

{F}rom all these results,  it would be tempting to say that it is the energy
landscape as a whole that becomes critical at point J.  Isostatic elasticity
would then be just one aspect of this criticality, another one could be the
intermediate power-law tail in the stress-drop distribution.  This critical
aspect is also illustrated by the intermittency in the stress response of
finite-size systems (see Fig.\ref{fig:stress_strain}) and by the growth of an
isostatic correlation length in the quasistatic response when point J is
approached from below \cite{heussingerPRL2009}.

At strain-rates above the quasistatic regime, the dynamics limits access to
certain regions of the energy landscape. While the dynamics is still highly
correlated, the dynamical correlation length, as measured by the amplitude of
the four-point susceptibility $\chi_4$, remains finite and actually
\emph{decreases} with lowering the volume-fraction towards $\phi_c$.


In the quasistatic regime we have shown that $\chi_4$ reflects, in two ways,
the interplay of elastic loading and plastic energy release
(Fig.~\ref{fig:chi4_amplitude}). First, the typical strain-scale of
heterogeneity is set by the frequency of plastic events. Second, the amplitude
of $\chi_4$ scales with system-size, which highlights the fact that the
quasistatic, plastic flow regime is, in fact, a finite-size dominated regime
with a correlation length that is limited by system size. This behavior should
be contrasted with the one observed below $\phi_c$, where a large but finite
correlation length has been identified, which is governed by the approach to
point J~\cite{heussinger10epl}.

Upon increasing the strain-rate we have shown that the correlation length starts
to decrease outside the finite-size scaling regime
(Fig.~\ref{fig:xi_dotgamma_phi}).  Olsson and Teitel~\cite{olssonPRL2007} infer
from their flow simulations that shear-stress should be viewed as a ``relevant
perturbation'' to point J, such that a different fixed-point and indeed
different physics is relevant for the flow behaviour at finite stress.  Our
findings support this picture for the dynamical correlations, which appear to
behave similarly to those observed in models of elasto-plastic flow
\cite{PicardEPJE2004,PicardPRE2005,bocquetPRL22009} or in low temperature
glasses \cite{tanguyEPJE2006,lemaitrePRL2009}: {correlations increase upon
  lowering the strain-rate and saturate at a system-size dependent value in the
  quasistatic regime.}

The flow behaviour in the vicinity of point J is therefore influenced by a
complex combination of two critical behaviour.  Large stress fluctuations
(relative to the yield stress), or geometrical changes (number of neighbours)
reflect the enhanced sensitivity of the material to small changes in external
conditions at point J, and are specific properties of the energy landscape at
this point. On the other hand, dynamical correlations above point J are
dominated by the system size, and build up progressively as the strain rate is
decreased, as in any elasto-plastic system, and are not particularly sensitive
to the proximity of point J.

{\bf Acknowledgments}
  The authors acknowledge fruitful discussions with Ludovic Berthier, Lyd\'eric
  Bocquet, Erwin Frey; Craig Maloney and Michel Tsamados, as well as thank the von-Humboldt
  Feodor-Lynen, the Marie-Curie Eurosim and the ANR Syscom program for financial
  support.

\bibliography{jamming}

\end{document}